\documentclass[graybox]{svmult}

\usepackage{mathptmx}       % selects Times Roman as basic font
\usepackage{helvet}         % selects Helvetica as sans-serif font
\usepackage{courier}        % selects Courier as typewriter font
\usepackage{type1cm}        % activate if the above 3 fonts are
\usepackage{makeidx}         % allows index generation
\usepackage{graphicx}        % standard LaTeX graphics tool
\usepackage{multicol}        % used for the two-column index
\usepackage[bottom]{footmisc}% places footnotes at page bottom

\usepackage{amsfonts}
\usepackage{amsmath}
\usepackage{amssymb}

\usepackage{tikz} 
\usetikzlibrary{arrows.meta}

\usepackage{subcaption}
\makeindex             % used for the subject index
\begin{document}

\title*{Experimental Design Issues in Big Data. The Question of Bias}
\author{Elena Pesce \and Eva Riccomagno \and Henry P. Wynn}
\institute{E. Pesce \at Department of Mathematics, University of Genoa, Italy, \email{pesce@dima.unige.it}
\and E. Riccomagno \at Department of Mathematics, University of Genoa, Italy, \email{riccomagno@dima.unige.it}
\and H.P. Wynn \at Department of Statistics, London School of Economics, UK, \email{h.wynn@lse.ac.uk}}
\maketitle

\abstract{Data can be collected in scientific studies via a controlled experiment or passive observation. Big data is often collected in a passive way, e.g. from social media. 
In studies of causation great efforts are made to guard against bias and hidden confounders or feedback which can destroy the identification of causation by corrupting or omitting counterfactuals (controls). Various  solutions of these problems are discussed, including randomization.}


\section{The challenges of experimental design with big data}
\label{sec:1}
The value of experimental design in physical and socio-medical fields is increasingly realised, but at the same time systems under consideration are
more complex. It may not be possible to do a carefully controlled experiment
in many areas, but at the same time huge quantities of data are being produced,
for example from social media and web-based transactions. An added problem is that the traditions of experimental design differ. For example in engineering design it will be possible to do a control experimental on a test bench, whereas in
the social-medical sciences the local counterfactual will be missing: we do not know how a particular patient would have fared if they were not given the drug.
Foundation work on these issues is by~\cite{b10}.
% Rosenbaum and Rubin (1983). 
Roughly, the causal effect can only be measured on the average, with great care taken about the background population, with more reluctance than in the physical sciences to extend the conclusions outside the population under study.  An old issue, which goes back into the history of science, is the distinction between active and passive observation. 
Is placing a sensor on a driverless car to collect data (for control) an intervention in the sense of the declaration that to prove causation you have to intervene?
  Despite these different historical traditions there seems to be general agreement (i) that deriving causal models is a kind of gold standard and (ii) that
to produce a causal model we need to guard against bias from different sources: hidden confounders, sampling bias, incomplete models, feedback and so on.
 
We cover a few of the ideas from the theory of causation (Section \ref{sec:3}) and then suggest that
the double activity of building causal models while at the same time guarding against bias has features of a cooperative game (Section \ref{sub:3.1}). At its simplest a randomized clinical trial is minimax solution to a game against the sources of bias. With this in mind we make the natural but speculative suggestion that we can import theories of Nash equilibrium and supply a simple example motivated by the theory of optimum experimental design under a heading of optimal bias design. We could have taken a Bayesian optimal design, for example from~\cite{b6,b12}. 
%Hainy et al (2014) or Sebastiani and Wynn (2000). 
But for this short paper we felt it was enough to allow our randomness to come from the error distribution or the randomization itself.

\section{Causal Models}
\label{sec:3}
A major critique of passive analysis of the machine-learning type is the lack of attention to the building of causal models. We discuss briefly the main ingredients of causal graphical models and then the implications for experimental design~\cite{b11}.

A causal model is often described via a direct acyclic graph (DAG), $G(E,V)$, where each vertex $i \in V$ holds a (possibly vector) random variable $X_i$. Care has to be taken with the edges $i \rightarrow j$. The natural intuition that the edge means $X_i$ causes $X_j$ is not correct, at least not without much qualification. The DAG is a vehicle for describing all conditional independence structures. 

We can define a variable $X_j$ which is never observed as \emph{latent}, also \emph{hidden}. There is a slight difference: hidden may be that we do not know it is there but it might be. Latent may also be taken as expressing prior information. Thus a latent layer in machine learning context may be included to allow a more complex model, such as a mixture model.

The conundrum with causal models stems from the distinction between passive observation and active experimental design. Experimental design is an \emph{intervention} and there are essentially two types. First, we can simply apply some kind of \emph{treatment} at node $i$ to obtain a special $X_i$, for example give a patient $i$ a drug. Second, and even more active, one can set variable $X_i$ to say high and low levels.

Passive observation means that a joint sampling distribution covers all observed $X_i$. The act of setting should be thought as advantageous in the sense that we are in some kind of classical or optimal design framework, but disadvantageous in that it is destructive. Roughly, setting $X_i$ destroys our ability to learn about the population from which $X_i$ comes. 

Consider a simple DAG: $X_1\longrightarrow X_2 \longrightarrow X_3 \longrightarrow X_4$ 
and for ease of explanation we write down a univariate linear version with obvious interpretation
\begin{align*}
X_1  &= \theta_0 + \epsilon_1       &X_2 =  \theta_1 X_1 + \epsilon_2 \\
X_3  &=  \theta_2 X_2 + \epsilon_3   &X_4 =  \theta_3 X_3 + \epsilon_4 
\end{align*}
where $\{\epsilon_i\}_{i=1,\ldots,4}$ are error variables.
Suppose we are interested in the last causal parameter $\theta_3$. 
Ideal would be to carry out a controlled experiment, setting the levels of $X_3$ and observing $X_4$. The first assumption to make is governed by the following: % \medskip

\emph{Principal 1.} The distribution of $X_4$ conditional on a \emph{set} value of $X_3$ is the same as when the same value of $X_3$ was passively observed. % \medskip

There are arguments to justify this but it remains a most important assumption. 
We can also passively observe $X_1,X_2,X_3,X_4$. Note that the model is \emph{nonlinear} in the parameters as 
$X_4 = \theta_0\theta_1\theta_2\theta_3 + \theta_1\theta_2\theta_3\epsilon_1+\theta_2\theta_3\epsilon_2+\theta_3\epsilon_3+\epsilon_4 
$ and also that $X_4$ is Gaussian if the $\{\epsilon_i\}$ are Gaussian. 
One may not have to choose between a controlled experiment and passive observation. This lead to another principal, see~\cite{b5} % \medskip

\emph{Principal 2.} A mixture of passive observation experiment and active experimentation may be optimal. % \medskip

There is considerable discussion in trying to understand how to learn for DAG models with interventions, and controlled experiments are a form of intervention. Most effort has been put into identifiability; see~\cite{b3} for a review. 
In our example suppose there is an extra arrow $X_1 \rightarrow X_4$. Such an arrow is referred to as a backdoor. If the index is time we can say that there is another path from $X_1$ into the future in addition to $X_1 \rightarrow X_2 \rightarrow X_3 \rightarrow X_4$. 
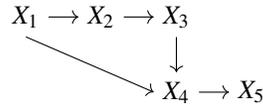
\begin{center}
\begin{tikzpicture}
\begin{scope}
    \node (X1) at (0,0) {$X_1$};
    \node (X2) at (1,0) {$X_2$};
    \node (X3) at (2,0) {$X_3$};
    \node (X4) at (2,-1) {$X_4$};
    \node (X5) at (3,-1) {$X_5$};

    \draw [->] (X1.east) -- (X2.west);
    \draw [->] (X2.east) -- (X3.west);
    \draw [->] (X1.south) -- (X4.west);
    \draw [->] (X3.south) -- (X4.north);
    \draw [->] (X4.east) -- (X5.west);
\end{scope}
\end{tikzpicture}
\end{center}
Now if we fix $X_3$ we cannot so simply estimate $\theta_3$ because the distribution of $X_4$ is corrupted by the new path. In the observational case, we have another parameter and the changed equation
\begin{equation}
\label{eq_example}
X_4 = \theta_3 X_3 + \theta_4 X_1 + \epsilon_4. 
\end{equation}
There are now too many parameters for the observations (even with replication).

The celebrated backdoor theorem due to ~\cite{b9} tells us how to obtain identifiability. Suppose you want to see whether $X_i$ causes $X_j$, then we need two conditions for a good conditioning set of variables $S$:
\begin{enumerate}
\item No node (variable) in $S$ is a descendent of $X_i$
\item $S$ blocks every (backdoor) path from $X_i$ to $X_j$ that has an arrow into $X_i$.
\end{enumerate}
This theorem tell us: (i) whether there is confounding given this DAG, (ii) if it is possible to remove the confounding and (iii) which variables to condition on to eliminate the confounding. For example, if we are trying to establish the effect of $X_3$ on $X_4$ then we must observe, or set and condition on, any $X_i$ which is not a descendant of $X_3$ and blocks all paths from, in our case, ancestors of $X_3$. In addition, if there are any other downstream (future) variable such as an extra $X_5$ with $X_4 \rightarrow X_5$, then $X_5$ will not interfere with our causal analysis; we can forget it. In summary % \medskip

\emph{Principal 3.} Guard against effects from nuisance confounders by suitable additional conditioning. % \medskip

\section{Bias Models}
\label{sec:4}

Before presenting our  contribution, we briefly review relevant literature. 
For the model without bias
\begin{equation} \label{nobiasmodel}
  \operatorname{E}\left[ Y_i \right] =  \theta^T f(x_i)   \end{equation}
   with $ \theta \in \mathbb R^p$, ${x_i} \in \mathcal X$ for $i=1,\ldots,n$, and the usual  assumptions on the random error, 
\cite{Drovandi} and~\cite{Stufken} propose information-based and sequential algorithms (also response adaptive in~\cite{Drovandi}) for the selection of a subsample from a large, or possibly big, dataset. They provide an optimal subsample with respect to a chosen utility function. 

Bias model and optimal design of experiments were considered by~\cite{b8} and recently in the context of big data by~\cite{Wiens}. 
Those authors add a bias term $ \delta^T h $ to the model~(\ref{nobiasmodel}) and thus study 
 $  \operatorname{E}\left[ Y_i \right] =
 \theta^T f(x_i) + \delta^T h(x_i) 
 $. 
  They search for a design which minimises the mean square error of the least square estimator of the $\theta$ parameters,  guarding $\theta$ from the bias term. 
 In particular~\cite{Wiens} proposes a theory of minimax $I$- and $D$-robust design as subset of a large finite set of points,
 while~\cite{b8} proves results for a design to be optimal when the effect of the bias term is bounded above from a given constant and below from zero.
 
The conditioning argument of the backdoor theorem is a way of avoiding biases. In the above example in Equation (\ref{eq_example}) $\theta_4$ gives a bias. Enough conditioning creates a kind of laboratory inside which we can conduct our experiment by setting the level of $X_3$. Sometimes this is referred to as creating a Markov blanket. But there are sources of bias which either we do not know at all or have some ideas about but are too costly to control. 
Biases range from those we really know about but simply do not observe to those which are introduced to model additional variability. This will affect the overall distribution of the observed variables, in a way similar to classical factor analysis. % \medskip

\emph{Principal 4.} Special models are needed to ascertain and guard against hidden sources of bias, for example using randomization or latent variable methods. % \medskip

We build on the ideas in~\cite{b8} and discuss in details how optimal experimental design can guard against hidden sources of bias, indicated below with the letter $z$. Thus consider a two part model in which the first part is the causal model of main interest with parameters $\theta$ and the second part is the bias term with parameters $\phi$. This separation is familiar from traditional experimental design where $\theta$ and $\phi$ might be treatment and block parameters, respectively,~\cite{b1,b8}. The model is:
\begin{equation}\label{eq1}
Y_i = \theta^T f(x_i) + \phi^T g(z_i) + \epsilon_i 
\end{equation}
where the $\epsilon_i$ are independent and have equal variance $\sigma^2$.

We want to protect the usual least square estimator, $\hat{\theta}$, 
obtained from the reduced model in Equation (\ref{nobiasmodel}) ignoring the bias term $\phi^T g(z_i)$.
Define the full moment matrix by
\begin{equation}
M =  \int (f(x)^T,g(z)^T)^T (f(x)^T,g(z)^T) \xi_{x,z}\; d(x,z)=
\left [ \begin{array}{cc}
M_{11} & M_{12} \\
M_{21} & M_{22} \end{array} \right], \nonumber
\end{equation}
where $\xi_{x,z}$ is the experimental design measure over $(x,z)$-space.
Then the mean squared error (MSE) matrix can be written as 
\[
 \mathbb{E} \{ (\hat{\theta} -\theta)(\hat{\theta} -\theta)^T \} = \sigma^2 N^{-1} R
\]
where
\begin{equation}
R = M_{11}^{-1} + M_{11}^{-1} M_{12} \psi \psi^T M_{21}M_{11}^{-1} = S_1+S_2 \nonumber
\end{equation}
with $\psi = \frac{N}{\sigma} \phi $ the standardised bias parameter and $N$ the sample size (see \cite{b8}).  

Well known criteria for optimality ask to minimise over the choice of experimental design the quantity:
$\mbox{trace} (R) = \mbox{trace} (S_1) + \mbox{trace} (S_2)
$ (the trace criteria or $A$-optimality)
or 
$\mbox{det} (R) = \mbox{det} (S_1) \; \left (1+    \psi^T M_{21} M_{11}^{-1} M_{12}\psi\right) 
$
(the $D$-optimality criteria).

The design problem is easier when the design space  and design $D$ are direct products and thus can be written as 
\begin{equation}\label{eq2}
{\mathcal{X}} \times {\mathcal{Z}},\;\; D = D_1 \times D_2 
\end{equation}
with $x\in {\mathcal{X}} $ and $z\in \mathcal{Z}$, $D_1$ and $D_2$ are finite subsets of, respectively, $\mathcal{X}$ and $\mathcal{Z}$. 
Then, $\operatorname{trace}(R)$ includes a term which depends only on $D_1$, likewise a factor in $\operatorname{det}(R)$ depends only on $D_1$.

The most familiar example is from clinical trials where one compares a treatment against a control. Consider the simple case
\begin{eqnarray*}
Y_{1i} &  = & \theta_1 + \theta_2  + \phi ( z_{1i}  - \bar{z})  + \epsilon_i \\
Y_{2j} &  = &  \theta_1-\theta_2  + \phi ( z_{2j} - \bar{z} )+ \epsilon_j ,
\end{eqnarray*}
where the $z_i$ are unwanted confounders which may be a source of bias, the $\bar{z}$ is the grand mean and 
$n=N/2$ points are allocated to each group.  
Adapting the above analysis we obtain
\begin{equation}
 M = \frac{X^\prime X}{N} = \displaystyle{ 
\left[
\begin{array}{ccc}
1 & 0 & 0 \\
0 & 1 &  (\bar{z}_1-\bar{z}_2)/2 \\
0 & (\bar{z}_1-\bar{z}_2)/2 & s \\
\end{array}
\right],
}
 \nonumber
\end{equation}
where the $\bar{z}_i$, $i=1,2$,  terms are the group means and $N s =\sum_{i = 1}^{n} (z_{1i} - \bar{z})^2 + \sum_{i = 1}^{n} (z_{2i} - \bar{z})^2$. 
The bias term is $\operatorname{trace}(S_2) = \psi^2 ( \bar{z}_1 - \bar{z}_2)^2/4$ which is  zero when $\bar{z}_1= \bar{z}_2$. 
This is the simplest case of  balance and extends easily to multivariate $z$. 
A number of methods of achieving balance have been studied, each of which can be cast in the above framework
\begin{enumerate}
\item Stratification: balancing in each stratum and then aggregating the difference.
\item Distance methods: pairing up treatment and control with which are close in {$z$-space} with respect to some distance such as Mahalanobis distance~\cite{b7}.
\item Propensity score. This much researched method seeks to balance in such way as to ensure that the bias correction is extended to a larger parent population~\cite{b10,b11}. Some adaptation of the above method analysis is possible in this case. 
\end{enumerate}

\subsection{A Game Theoretic Approach}
\label{sub:3.1}
For ease of explanation we introduce two players: Alice (A) and Bob (B).
Alice selects a  causal model design  $D_1$ using $\{\theta, f\}$ and Bob selects  design $D_2$ using $\{\phi, g\}$. In the product case (\ref{eq2}), Alice and Bob can operate separately. In other cases they may cooperate fully to find the best design over the design space for the pair $(x,z)$. However there is another  possibility, namely to use a Nash equilibrium approach~\cite{b2,b4,b13}.

For two players A and B with composite cost functions $C_1(u,v), C_2(u,v)$ and solutions $u^*,v^*$ at equilibrium it holds
\begin{equation}
\begin{array}{cccc}
\mbox{Alice}:  &  u^* & = & \underset{u}{\operatorname{argmin }} \,\, C_1(u, v^*)\\
\mbox{Bob}:   &  v^*  & = & \underset{v}{\operatorname{argmin }} \,\,  C_2 (u^*, v)
\end{array} \nonumber
\end{equation}
We illustrate the presence of Nash equilibrium in causation-bias set up by a simple example. 
We take a distorted design space, but still a product-type design measure. Thus, let the model be
\[
\mathbb{E}(Y) = \theta_0 + \theta_1 x + \phi z 
\]
and let the design have a four support points (we put the design measure in the second line):
\begin{equation}
\left\{
\begin{array}{cccc}
(1,1), & (0,1), & (0,-1), & (-1,-1) \\
\alpha \beta, &\;\; (1-\alpha)\beta, &\;\;\alpha (1-\beta),&\;\; (1-\alpha)(1-\beta)
\end{array}
\right\}, \nonumber
\end{equation}
where $ 0  \leq \alpha,\beta \leq 1$

Since, in this case, $M_{12}$ is a $2 \times 1$ column vector:
\begin{equation}
\mbox{trace}(S_2)  =  \psi^2 M_{21} M_{11}^{-2} M_{12} \nonumber
\end{equation}
The equilibrium takes the form:
\begin{equation}
\begin{array}{cccc}
\mbox{Alice}:  &  \alpha^* & = & \arg \min_{\alpha} \mbox{trace}(S_1) \\
\mbox{Bob}:   &  \beta^*  & = & \arg \min_{\beta} \mbox{trace}(S_2)
\end{array} \nonumber
\end{equation}
There are two Nash equilibria given by solving 
%\begin{equation*}
$\frac{\partial}{\partial \alpha} \mbox{trace}(S_1) = \frac{\partial}{\partial \beta} \mbox{trace}(S_2) = 0. $
%\end{equation*}
This gives two solutions $(\alpha^*,\beta^*)$ and $(\frac{1}{2}, \frac{1}{2})$ with  $\alpha^* = 0.59306$ and $\beta^* = 0.08274$ computed numerically.  Note that both solutions do not depend on $\psi$, and in fact scale invariance of this kind is a well known feature of Nash equilibrium.

We can compare the solution with an overall optimisation by setting $\psi=1$ and minimizing  $\mbox{trace}(S_1) + \mbox{trace}(S_2)$. The minimum is $4$, it is achieved at $(\alpha,\beta) = (\frac{1}{2}, \frac{1}{2})$ with $(\mbox{trace}(S_1), \mbox{trace}(S_2)) {=} (3 , 1)$. Whereas at $(\alpha^*,\beta^*)$ the value of $\mbox{trace}(S_1) + \mbox{trace}(S_2)$ is approximated to $5.1735$ with $(\mbox{trace}(S_1), \mbox{trace}(S_2)) {=} (4.483,0.6905)$.

Let us return to the role of Bob in our narrative. His experimental design decision will depend on his knowledge about the
bias. For ease of explanation we reduce the argument to two canonical cases. % \medskip

Approach 1. Unknown $\psi$
\begin{equation}
\mbox{trace}(S_2) = \mbox{trace} \left(M_{11}^{-1} M_{12} \psi \psi^T M_{21} M_{11}^{-1}\right) = \psi^T Q _1\psi \nonumber
\end{equation}
\begin{equation}
Q_1= M_{21}M_{11}^{-2} M_{12}. \nonumber
\end{equation}
Under a restriction $||\psi ||=1$ this achieves a maximum at the maximum eigenvalue: $\lambda_{\max} (Q_1)$. We can take this as our criterion which is close to the $E$-optimality of optimum design theory. % \medskip

Approach 2. In equation (\ref{eq1}), for unknown $\phi^T g(z) = h(z) \in {\mathcal H}$ in some function class, we have
\begin{equation}
||\mbox{E} (\hat{\theta}) - \theta||^2 = h(z)^T Q_2 \; h(z) \nonumber
\end{equation}
\begin{equation}
Q_2= X_1 M_{11}^{-2} X_1^T \nonumber
\end{equation}
where $X_1=[f(x)]_{x\in D_1}$. 
We cannot optimise over $x$ $(X_1)$ because, in our narrative, Alice needs it for the causal parameter $\theta$.  A solution is then 
\begin{equation}
\min_{P_z} \;\mathbb{E}_Z \left\{\max_{h \in {\mathcal H}} \left(h(z)^T Q_2 \; h(z)\right) \right\}, \nonumber
\end{equation}
where $P_z$ is the randomization distribution. In the language of game theory this is a mixed strategy to achieve a minimax solution. 

Randomisation has been heralded as one of the most important contributions of statistics to scientific discovery. There are several arguments put forward for using randomization: (i) it helps support assumptions of exchangeability in a Bayesian analysis (ii) it supports classical zero mean and equal variance arguments and (iii) it produces roughly balanced samples. 

\section{Conclusion}
\label{sec:conclusion}

After a discussion of some issues related to the use of experimental design to help establish causation in complex models, we study in a little more detail the use of optimal design methods to remove bias. In the standard case the causal part of a model can be estimated orthogonally from the bias. In more complex cases the problem can be set up as a co-operative game.  We demonstrate the existence of Nash equilibria for a small example and point to a formulation which would include randomization. This is preliminary work, establishing  model classes (for example special $h$'s, $\mathcal{H}$'s, $P_z$'s) and conditions on $D$  for which Approaches~1 and~2 can be turned into efficient algorithms is object of current work. The general proposition is that such methods will help protect causal models against bias.

\begin{acknowledgement}
We thank the anonymous reviewers for thorough reading of the manuscript. 
\end{acknowledgement}
 
%%%%%%%%%%%%%%%%%%%%%%%% referenc.tex %%%%%%%%%%%%%%%%%%%%%%%%%%%%%%
% sample references
% %
% Use this file as a template for your own input.
%
%%%%%%%%%%%%%%%%%%%%%%%% Springer-Verlag %%%%%%%%%%%%%%%%%%%%%%%%%%
%
% BibTeX users please use
% \bibliographystyle{}
% \bibliography{}

\begin{thebibliography}{99.}

\bibitem{b1} Box, G.E., Draper, N.R.: A basis for the selection of a response surface design. \textit{Journal of the American Statistical Association}, 54(287), 622--654 (1959)

\bibitem{b2} Cheng, C.S., Li, K.C.: A minimax approach to sample surveys. \textit{The Annals of Statistics}, 552--563 (1983)

\bibitem{Drovandi} Drovandi, C.C., Holmes, C., McGree, J.M., Mengersen, K., Richardson, S., Ryan, E.G.: Principles of Experimental Design for Big Data Analysis. \textit{Statistical Science}, 32(3), 385--404  (2017)

\bibitem{b3} Drton, M., Weihs, L.: Generic identifiability of linear structural equation models by ancestor decomposition. \textit{Scandinavian Journal of Statistics}, 43(4), 1035--1045 (2016)

\bibitem{b4} Grant, W.C., Anstrom, K.J.: Minimizing selection bias in randomized trials: A Nash equilibrium approach to optimal randomization. \textit{Journal of Economic Behavior \& Organization}, 66(3), 606--624 (2008)

\bibitem{b5} Hainy, M., M{\"u}ller, W.G., Wynn, H.P.: Approximate Bayesian computation design (ABCD), an introduction. \textit{In: mODa 10–Advances in Model-Oriented Design and Analysis}, 135--143. Springer, Heidelberg (2013)

\bibitem{b6} Hainy, M., M{\"u}ller, W.G., Wynn, H.P.: Learning functions and approximate Bayesian computation design: ABCD. \textit{Entropy}, 16(8), 4353--4374 (2014)

\bibitem{b7} LaLonde, R.J.: Evaluating the econometric evaluations of training programs with experimental data. \textit{The American economic review}, 604--620 (1986)

\bibitem{b8} Montepiedra, G., Fedorov, V.V.: Minimum bias designs with constraints. \textit{Journal of Statistical Planning and Inference}, 63(1), 97--111 (1997)

\bibitem{b9} Pearl, J.: \textit{Causality}. Cambridge university press (2009) 

\bibitem{b10} Rosenbaum, P.R., Rubin, D.B.: The central role of the propensity score in observational studies for causal effects. \textit{Biometrika}, 70(1), 41--55 (1983)

\bibitem{b11} Rubin, D.B.: Bayesian inference for causal effects: The role of randomization. \textit{The Annals of Statistics}, 34--58 (1978)

\bibitem{b12} Sebastiani, P., Wynn, H.P.: Maximum entropy sampling and optimal Bayesian experimental design. \textit{Journal of the Royal Statistical Society: Series B}, 62(1), 145--157 (2000)

\bibitem{b13} Stenger, H.: A minimax approach to randomization and estimation in survey sampling. \textit{The Annals of Statistics}, 395--399 (1979)

%\bibitem{b14} Stigler, S.M.: The use of random allocation for the control of selection bias. \textit{Biometrika}, 56(3), 553--560 (1969)

\bibitem{Stufken} Wang, H., Yang M., Stufken, J.: Information-Based Optimal Subdata Selection for Big Data Linear Regression. \textit{Journal of the American Statistical Association} (in press)

\bibitem{Wiens}  Wiens, D.P.: I-robust and D-robust designs on a finite design space. \textit{Statistics and Computing}, 28(2), 241--258 (2018) 

\end{thebibliography}
%

\end{document}